\documentclass[aps,prb,showpacs,superscriptaddress,twocolumn]{revtex4-1}
\usepackage{subfigure}
\usepackage{float}
\usepackage{bm}
\usepackage{amsmath,amsfonts,amssymb,graphicx}
\begin{document}
\raggedbottom
\title{Composite Pulses in N-level Systems with SU(2) Symmetry and their Geometrical Representation on the Majorana Sphere}
\author{H. Greener}
\author{H. Suchowski}
\email{haimsu@post.tau.ac.il}
\affiliation{School of Physics and Astronomy, Tel Aviv University, Tel Aviv 69978}ֿ
\affiliation{Center for Light-Matter Interaction, Tel Aviv University, Tel Aviv 69978}

\begin{abstract}
High fidelity and robustness in population inversion is very desirable for many quantum control applications. We expand composite pulse schemes developed for two-level dynamics, and present an analytic solution for the coherent evolution of an N-level quantum system with SU(2) symmetry, for achieving high fidelity and robust population inversion, which outperforms common solutions in N-level dynamics. Our approach offers a platform for accurate steering of the population transfer in physical multi-level systems, which is crucial for fidelity in quantum computation and achieving fundamental excitations in nuclear magnetic resonances and atomic physics. We also introduce and discuss the geometrical trajectories of these dynamics on the Majorana sphere as an interpretation, allowing to gain physical insight on the dynamics of many-body or high-dimensional quantum systems.
\end{abstract}

\pacs{32.80.Qk, 42.50.Dv, 42.65.Re, 82.56.Jn}

\maketitle

\section{Introduction}
Complete population transfer (CPT) from one quantum state to another has been the focus of extensive research these past few decades. This is generally achieved by shaping the duration and area of an electromagnetic pulse impinged on a system in order to excite it, and by employing various techniques ensuring this excitation is robust to inaccuracies in the system parameters and the pulse shape itself. Such control is desired for  obtaining high fidelity in quantum computation and quantum information processing \cite{QuantumInfo1, QuantumInfo2}, coherent manipulation of population inversion in atomic and molecular quantum systems \cite{AllenEberly, IntroAtomicMol2, IntroAtomicMol3, PythagCoupling}, directional optical waveguides \cite{DirectionalWGs1} and spin control in nuclear magnetic resonances\cite{KeelerNMR}. 

A handful of solvable models have been suggested and widely used for the coherently induced dynamics of two-level quantum systems. These include Rabi oscillations, in which a constant external field oscillating near the resonance of a two-level system is applied to achieve population inversion \cite{RabiOsc} due to an exact odd number of so-called $\pi$ pulses. This solution is very sensitive to experimental constraints, such as a mismatch beween the frequency of the external field and the system's resonance, known as detuning. Thus, other time dependent methods were derived, including the Landau-Zener, Rosen-Zener and Allen-Eberly models \cite{Zener,RosenZener,AllenEberly}, which allow for an adiabatic solution for very robust population inversion of a two-level system. However, these examples require very long and precise manipulation of the system and excitation parameters, which are not always feasible under experimental consequences. 

Another class of coherent solutions are composite pulses, which overcome such experimental constraints, and relax the need for a perfect system and excitation mechanism. These are a sequence of pulses with specifically chosen phases, commonly used in nuclear magnetic resonance (NMR), and for broadband population inversion by ultrashort pulses \cite{CP1,CarrSpinEcho,CP2,CP3,CP4,CP5,CP6,CP7}. Since NMR spectroscopy requires precise pulse excitation for spin population inversion, attention was turned to designing composite sequences to compensate for conventional single pulse imperfections \cite{LevittLongArticle}. These may be due to spatial inhomogeneity, resonance offset and bandwidth. Thus the performance of composite pulses in guiding a system to a final state is feasible, accurate and robust. 

Among these pulse sequences are Levitt's widely used spin population inversion schemes for NMR excitation\cite{CP4}. This composite pulse enables a two-level spin $j=\frac{1}{2}$ system to undergo accurate excitation, regardless of the pulse or system's imperfections, by steering the system step-wise through three pulses. This scheme has since opened a wide variety of different composite pulse sequences, that revolutionized the field of NMR and its applications. While Levitt's composite solution, which is comprised of pulses with rectangular temporal shape, is suitable for NMR experiments, it fails to maintain its efficient nature for pulses of ultrashort timescales. Thus a different composite pulse scheme was suggested recently by Torosov \emph{et al.} \cite{VitanovSmooth, VitanovUltraBB} for pulse envelopes of smooth temporal shape , such as Gaussian pulses. Using these solutions, one can accurately excite two-level optical systems by tailoring the phases of a composite ultrashort pulse sequence regardless of the exact ingredient pulse shapes. 

Despite the success of the above schemes for two-level systems, studies of the multi-level case have remained sparse in the field of NMR \cite{Levitt3Levels, LevittLongArticle}, atomic and optical systems. Several schemes have been suggested for the evoloution of multi-level systems. Solutions such as adiabatic elimination, Electromagnetically-Induced-Transparency,  stimulated Raman adiabatic passage (STIRAP) and the Landau-Zener picture \cite{PhysRevA.40.6741, PhysRevA.29.690, RevModPhys.70.1003, Shore1992, EIT, STIRAPreview} have been thoroughly studied to achieve the controlled evolution of a three-level system to a chosen final state. Additionally, Cook and Shore and Hioe \cite{PhysRevA.20.539, Hioe:87} introduced a method of exciting an N-level system with SU(2) symmetry, to achieve an extension of Rabi oscillations between palindromic states.  Yet all of these require the fine-tuning of experimental parameters in order to achieve efficient population inversion between the ground state and the excited state

Here, we expand the composite pulse schemes  developed for NMR and ultrashort temporal pulses to the general case of an N-level system. Such composite pulses were derived for three-level systems in the past \cite{Levitt3Levels}, and now we introduce a generalization of the NMR spin inversion scheme to a spin $j$ system, displaying SU(2) symmetry. In the case of a magnetically excited two-level system, the spacings between the levels are known to be equal. This is not the general case in atomic systems, thus we can apply these schemes to optical systems. Particularly, we show that we can achieve robust population inversion via ultrashort pulses between the first and the $Nth$ level of an N-level system with SU(2) symmetry. This is achieved by employing the composite sequences derived by Torosov \emph{et al.} \cite{VitanovSmooth}, and particularly broadband (BB), narrowband (NB) and passband (PB) excitation schemes. Furthermore, our approach allows for precise control of the above processes in population inversion between palindromic states, namely between levels $m$ and $N-m+1$. These enable the controlled manipulation of the dynamics of excitation processes in multi-level materials, for the design of high fidelity infrastructure for quantum information. 

Last, we show that the evolution of the solutions to the composite pulse schemes can be interpreted and visualized as trajectories on a Majorana sphere\cite{MajoRep, RevModPhys.17.237}, the only means available today to compactly visualize N-level dynamics with such symmetry on a unit sphere. This representation provides a useful tool to gain insight on high-dimensional or many-body systems, such as quantum entanglement in multiqubit systems \cite{NQubitStates, PhysRevA.94.022123}, spinor boson gases \cite{PhysRevB.80.024420,PhysRevA.85.051606}, geometric phases in many-body systems \cite{PhysRevLett.108.240402, PhysRevLett.99.050402, PhysRevE.78.021106}. This representation allows to feasibly portray the temporal evolution of many-level dynamics.

The paper is organized as follows. In Sec. II, we introduce pulse-induced coherent dynamics of a multi-level system and formulate NMR and ultrashort pulse sequences for N-level systems. In Sec. III, we employ the Majorana representation to geometrically depict these dynamics. Sec. IV provides a brief discussion and summary.

\section{Composite Pulse Schemes for Population Inversion in N-level Systems}
\begin{figure}
\includegraphics[width=3.0in]{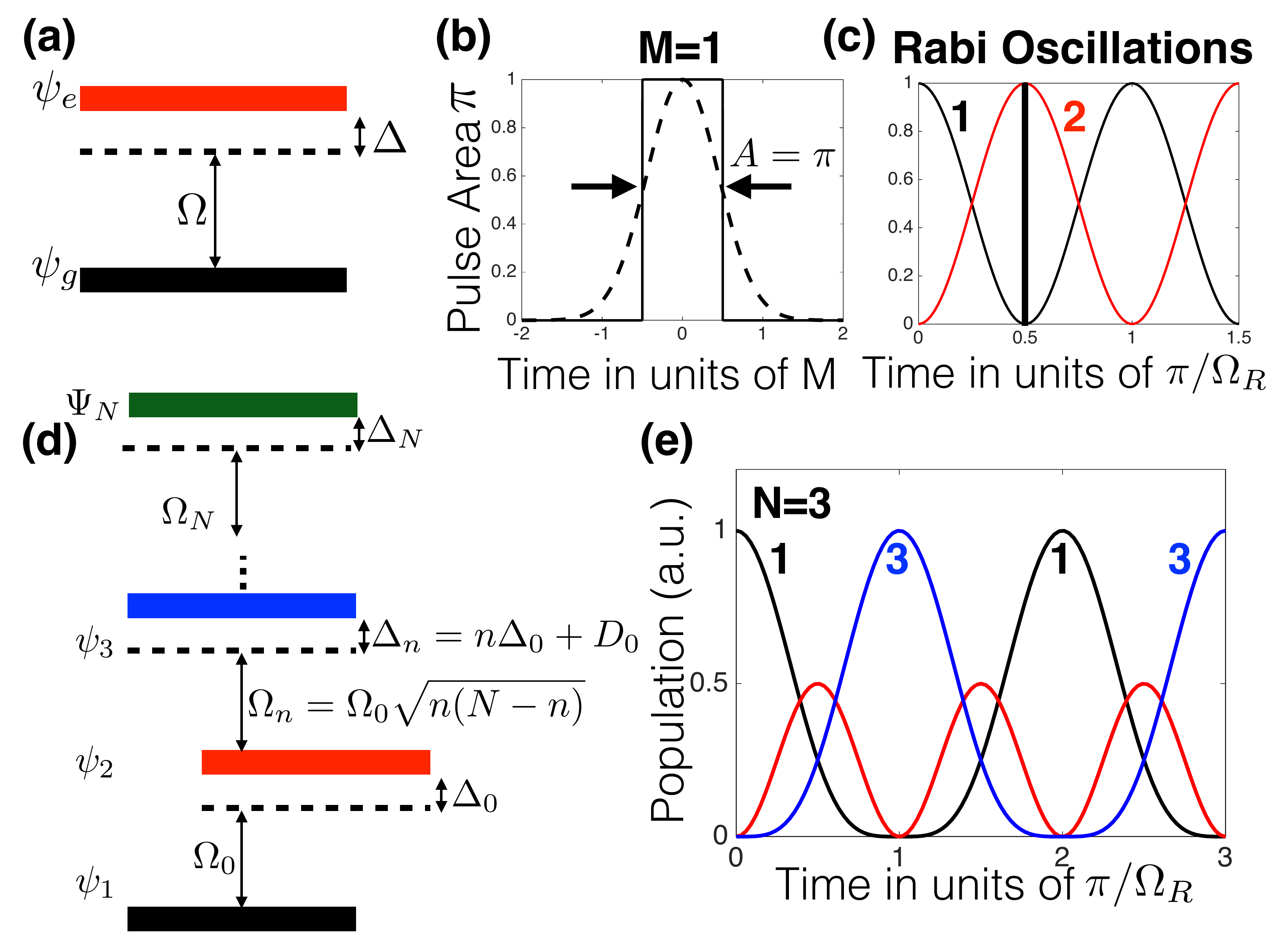}\centering
\caption{(Color online)\textbf{ Coupling scheme in atomic physics.} (a) A general
two-level coupled system, with a ground $\psi_{g}$ and excited $\psi_{e}$ level, coupling between adjacent levels
$\Omega$ and detuning $\Delta$.
(b) Continuous line: A single ($M=1$) pulse of rectangular temporal shape, defined as a $\pi$ pulse. Dashed line: A single $\pi$ pulse of Gaussian temporal shape. (c) The population evolution of a two-level system for the initial condition
$\psi_{1}=(1,0)$ under a single $\pi$ pulse. Population inversion is marked by a black dashed line at $t=j\pi/\Omega_{R}$, where for a two-level system, $j=0.5$. (d) General
N-level coupled system, with a ground $\psi_{1}$ and excited $\psi_{N}$ level, coupling between adjacent levels $\Omega_{n}$ and detunings $\Delta_{n}$. (e) The population evolution of a three-level system for the initial condition
$\psi_{1}=(1,0)$. Population inversion occurs at $t=j\pi/\Omega_{R}$, where for a three-level system $j=1$. \label{fig:Fig1}}
\end{figure}

We start by describing the coherent dynamics of an N-level system is described by a time
dependent Schrodinger equation for the probability amplitude for the
$n$th level\cite{ShoreBook,AllenEberly,PhysRevA.20.539}:
\begin{eqnarray}
i\frac{d}{dt}\psi_{n}(t) & = & \sum_{n'=1}^{N}H_{nn'}(t)\psi_{n'}(t)
\label{Schrodinger}
\end{eqnarray}

In the case where N=2, shown in Figure \ref{fig:Fig1}(a), the Hamiltonian in the rotating-wave-approximation (RWA), in which the applied field intensity is low and near resonance, is:
\begin{eqnarray}
H & = & \frac{\hbar}{2}\left(\begin{array}{cc}
0 & \Omega(t)e^{-iD(t)}\\
\Omega^{*}(t)e^{iD(t)} & 0
\end{array}\right)
\label{RWAHam}
\end{eqnarray}
where  $\Omega(t)$ is the Rabi frequency for electric dipole transitions and the detuning between the laser carrier frequency and the energy gap in a two-level system is described by $D(t)=\int_{t_{i}}^{t}\Delta(t')dt'$, $\Delta=\omega_{0}-\omega$, in which $\omega$ denotes the laser carrier frequency and $\omega_{0}=(E_{2}-E_{1})/\hbar$ is the Bohr transition frequency between the two levels. 

The propagator describing the SU(2) dynamics of the two-level system is described by $\hat{U}=\int_{t_{i}}^{t}exp{[iH(t')t']}dt'$, such that $\psi(t_{f})=\hat{U}\psi(t_{i})$. Thus, for the special case of resonant lossless excitation, i.e. $\Delta=0$, one could define the area of a pulse as $\int_{t_{0}}^{t_{f}}\Omega(t')dt'$. A single rectangular shaped pulse of area $A=\pi$, shown as the continuous line in Figure \ref{fig:Fig1}(b), will lead to the commonly-known Rabi oscillations between the two levels, as shown in Figure \ref{fig:Fig1}(c). 
The propagator describing the evolution of the above system excited by a sequence of M pulses with areas $A_{t_{k}}$ and phases $\phi_{k}$ is given by the product $U_{M}=U_{\phi_{t_{f}}}(A_{t_{f}})...U_{\phi_{t_{i}}}(A_{t_{i}})$ (see Appendix A).

We first show that the description of a propagator specifying $M$ pulses can immediately be extended to an N-level system with SU(2) dynamics. In this case, the Hamiltonian of an N-level system lies entirely in the subspace spanned by generators of the SU(2) group, and can be written as $H(t)=c_{1}(t)\hat{\sigma}_{x}+c_{2}(t)\hat{\sigma}_{y}+c_{3}(t)\hat{\sigma}_{z}+c_{0}\hat{I}$,
where $\hat{\sigma}_{i}$ are the angular-momentum operators of spin
$j=\frac{1}{2}\left(N-1\right)$, $c_{i}(t)$ are arbitrary functions
of time and $\hat{I}$ is the identity matrix. In the case of interest where the RWA is applied and the couplings are non-vanishing only between adjacent levels, we can write $c_{1}(t)=Re\left\{ \Omega(t)\right\} $, $c_{2}(t)=Im\left\{ \Omega(t)\right\}$, $c_{3}(t)=\Delta(t)$ and $c_{0}(t)=0$. Thus for a sequence of M pulses with a constant area $A$ and phases $\phi_{k}$, where $k=1,...,M$, the Rabi frequency for step $k$ is $\Omega_{k}(t)=Ae^{-i\phi_{k}}$, the system evolves into a final state as $\psi(t_{f})=e^{iH_{M}t}...e^{iH_{1}t}\psi(t_{0})$ where
\begin{eqnarray}
H_{k}(t) & = & Re\left\{ \Omega_{k}(t)\right\} \hat{\sigma}_{x}+Im\left\{ \Omega_{k}(t)\right\} \hat{\sigma}_{y} + \Delta(t)\label{eq:hamiltonian}
\end{eqnarray}
This describes the irreducible representation of an N-level system with SU(2) symmetry. 

One of the known solutions to N-level coherent dynamics was suggested by Cook and Shore \cite{PhysRevA.20.539} as a generalization of the Rabi solution, and further developed by Hioe for other schemes of systems with SU(2) dynamics \cite{Hioe:87}. This solution considers time independent couplings and detuning parameters between adjacent levels, one could still apply specific pulses which result in Rabi oscillations. Figure \ref{fig:Fig1}(d) depicts an example of a general N-level coupled system
$\psi_{1},\psi_{2},...,\psi_{N}$, which is dictated by a sequence of Rabi frequencies $\Omega_{n}=\Omega_{0}\sqrt{n(N-n)}$ where $\Omega_{0}$ is the coupling between the first two levels which is a complex scaling factor, and the detunings $\Delta_{n}=n\Delta_{0}+D_{0}$, where $D_{0}$ is an arbitrary real number \cite{PhysRevA.20.539}.  

We recall that in order to achieve population inversion in a two-level system, one must apply a so-called $\pi$ pulse on the system. Namely, one must ensure that $\frac{\theta}{2} = \Omega_{R}t$ where $\theta = \pi$. We would like to generalize this to the N-level case, where by applying a single $\pi$ pulse, the population will flow from level 1 to level N in a characteristic time we refer to as the "coherence time" $\tau=\pi/\Omega_{R}$. Figure \ref{fig:Fig1}(e) is an example of such Rabi oscillations for an $N=3$ level system. Note a twofold increase in the typical time for population inversion between the two and three level systems. This is consistent with the fact that the coherence time for population inversion is proportional to the respective spin value, $\tau=j\pi/\Omega_{R}$ multiplied by the number of pulses M impinged on the system, which in the simplest case, $M=1$. 

\begin{figure*}
\begin{minipage}{\columnwidth}
\includegraphics[width=3.0in]{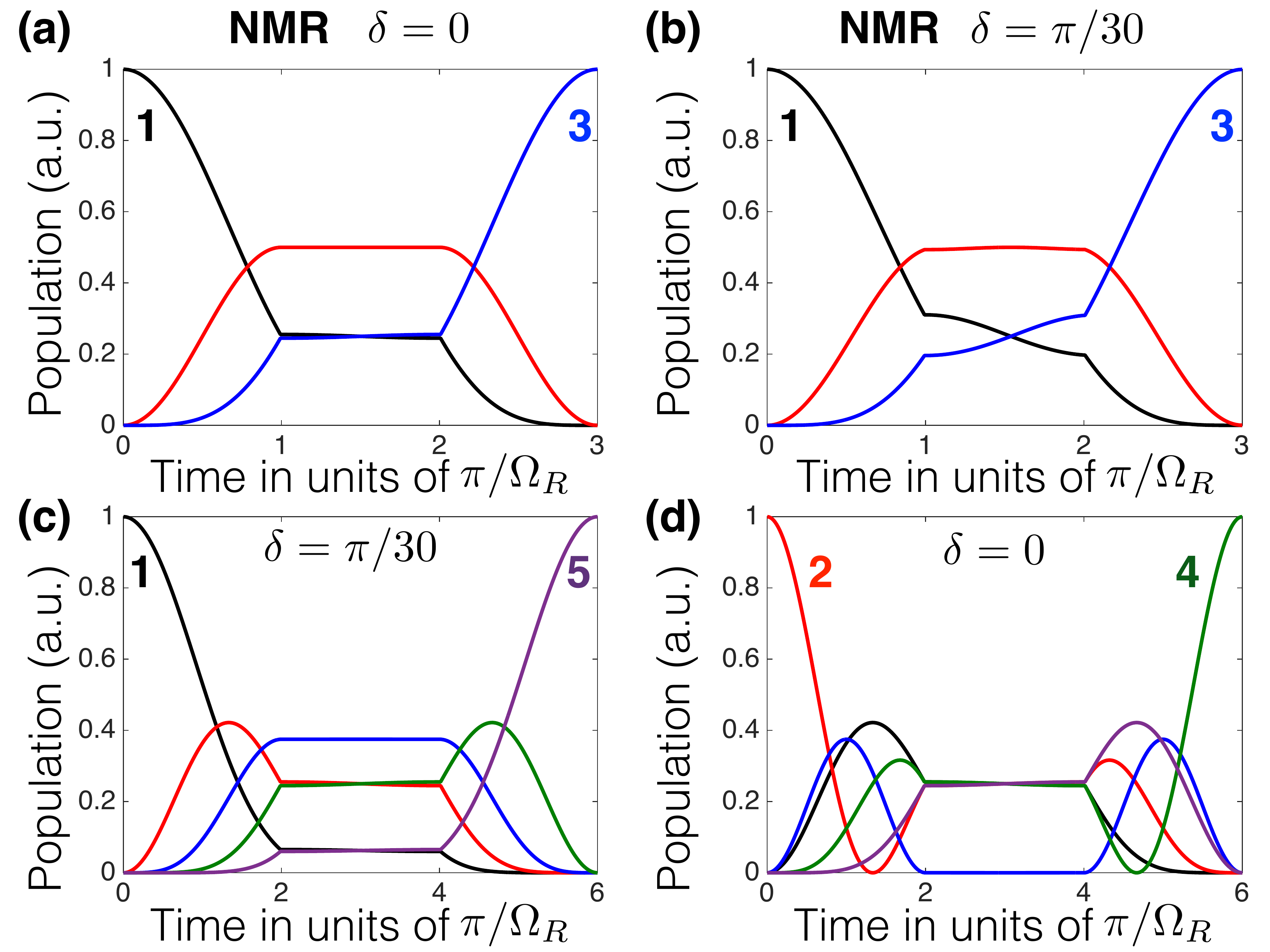}\centering
\end{minipage}
\begin{minipage}{\columnwidth}
\includegraphics[width=3.0in]{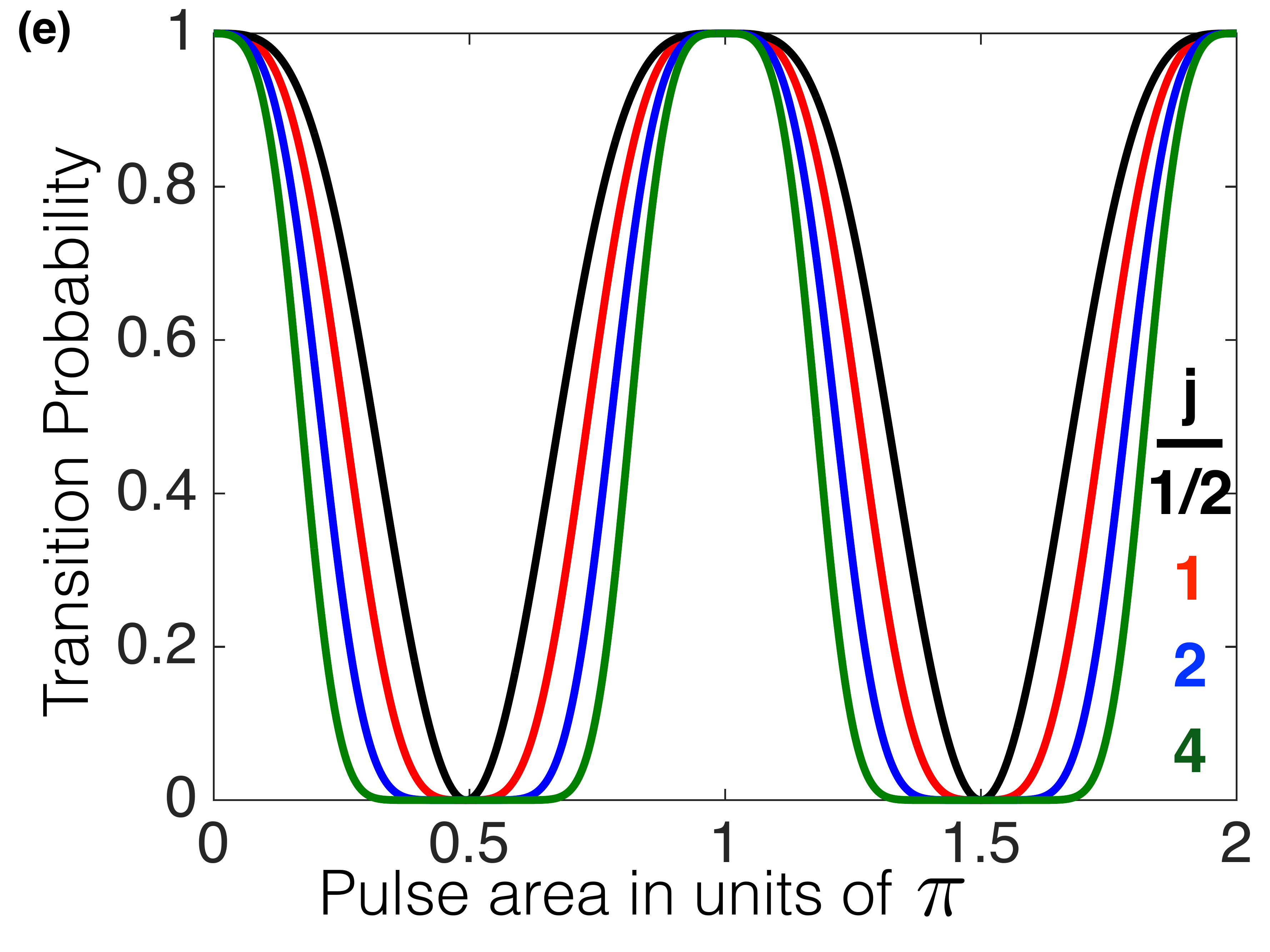}\centering
\end{minipage}
\caption{(Color online) \textbf{NMR population inversion scheme in N-level systems.} (a) Population inversion plot for three-level spin $j=\frac{1}{2}$ system with excited by a completely accurate $\pi$ pulse ($\delta=0$). (b) Same system with $\delta=\pi/30$. (c) Intensity plot for five-level spin $j=2$ system with $\delta=\pi/30$. (d) Population inversion of palindromic levels. Here, $\Psi(0)=\Psi_{2}$ with $\delta=0$. (e) Transition probability for full spin population inversion as a function of deviations from the pulse area $\delta$ in units of $\pi$ for various spin $j$ systems. Over 99\%  fidelity is achieved for pulse area inaccuracies of $A=\pi+\delta$, where $\delta=\pi/30$, in the case of a two-level and five-level system (see text for details). \label{fig:FigNMR1}}
\end{figure*}

However, a single pulse does not account for imperfections of a quantum system, due to spatial inhomogeneity or a resonance offset. Additionally, an inaccurate temporal shaped single pulse will not suffice to excite such a system, thus a composite pulse sequence for multilevel systems is necessary. Such problems were at the origin of the emergence of the composite pulse as a possible solution.

The field of nuclear magnetic resonance spectroscopy has traditionally
been considered the first to popularize composite pulses.  Spin echo pulse
sequences that compensate for static field inhomogeniety, were initially
used to achieve spin population inversion via radiofrequency (rf)
pulses. This scheme was further expanded and improved by Levitt \emph{et al.} \cite{CP4} to compensate for inhomogeneous single pulses and detuning effects of a spin $\frac{1}{2}$ system. Thus, three ingredient rf pulses of arbitrary inaccurate shape performed consecutively on a two-level spin $\frac{1}{2}$ system, result in a robust population inversion. This is known as the NMR spin population inversion composite scheme.

We generalize these pulse schemes to achieve robust and accurate population inversion in an N-leve system with SU(2) symmetry.  Furthermore, we show that our solution allows for population transfer between palindromic levels, namely to couple between levels $m \longleftrightarrow m'=N-m+1$.

We start by generalizing Levitt's \cite{CP4} NMR spin $\frac{1}{2}$ population inversion composite pulse to a three-level spin $j=1$ system. In this case, we choose:
\begin{eqnarray}
\begin{array}{ccc}
A_{\tau_{1}}=\frac{\pi}{2} &  & \phi_{\tau_{1}}=0  \\
A_{\tau_{2}}=\pi &  & \phi_{\tau_{2}}=\frac{\pi}{2} \\
A_{\tau_{3}}=\frac{\pi}{2} &  & \phi_{\tau_{3}}=0
\end{array}
\end{eqnarray}

We find that the inversion is between the first and third states, and each pulse should be applied for a duration of $\tau_{i}=\pi/\Omega_{R}$, as seen in Figure \ref{fig:FigNMR1}(a), such that the total coherence time for complete population inversion is $\tau=3\pi/\Omega_{R}$. This is consistent with the notion that this total coherence time should be proportional to $M$, the total number of pulses applied in the composite scheme.

Figures \ref{fig:FigNMR1}(b) and (c) are examples of complete population inversion in three-level spin $j=1$ and five-level spin $j=2$ systems, compensated by the above composite sequence for pulse area inaccuracies of $A = \pi + \delta$, where $\delta=\pi/30$. Over $99\%$ fidelity is achieved in both cases. To compare, for $\delta=\pi/10$, over $96\%$ fidelity is measured at the expected time of complete population inversion. A few examples of transition probability profiles for full spin population inversion as a function of deviations from the pulse area in units of $\pi$ are shown in Figure \ref{fig:FigNMR1}(e) for two-level spin $j=\frac{1}{2}$ (black), three-level $j=1$ (red), five-level $j=2$ (blue) and nine-level $j=4$ (green) systems.

Furthermore, by varying the initial conditions of the system, one could achieve robust spin inversion between any two palindromic states. To demonstrate this, we calculated the dynamics of a five-level spin $j=2$ system exposed to the NMR spin inversion composite pulse, set at $\Psi(t=0)=\Psi_{2}$. In Figure \ref{fig:FigNMR1}(d), one sees that an efficient inversion occurs between the 2nd and 4th states, along a coherence time of $\tau=6\pi/\Omega_{R}$, equivalent to the coherence time of the same system set at the initial condition of $\Psi(0)=\Psi_1$, seen in Figure \ref{fig:FigNMR1}(c). 

While Levitt's composite scheme has made a huge impact in NMR, and consequentially in MRI, its nature is inadequate for the excitation of dynamics in other two-level systems. An example of these include optical systems, controlled by ultrashort pulses, which are more prone to inaccuracies in pulse area, due to their smooth temporal shape. A comparison of a rectangular radiofrequency and a Gaussian ultrashort temporal pulse shape is shown in Figure \ref{fig:Fig1}(b). The rectangular temporal shaped pulse can accurately be designed with a pulse are of $\pi$, while for the ultrashort Gaussian pulse, this conveys a challenge. For this reason, Torosov \emph{et a.} \cite{VitanovSmooth,VitanovUltraBB} designed a series of composite pulse sequences to accurately achieve an arbitrarily flat inversion profile by tailoring the phases of the ingredient pulses. Thus, a broadband (BB) pulse scheme \cite{VitanovSmooth} for a two-level system is achieved by introducing such a composite
sequence in which a flat top of the excitation profile is required
at a pulse area of $A=\pi$. Alternatively, one could require a flat
bottom of the excitation profile at pulse area $A=0$ in order to
achieve narrowband (NB) pulses \cite{VitanovSmooth}, or both a flat top at $A=\pi$ \emph{and}
a flat bottom at $A=0$ to achieve a passband (PB) pulse \cite{VitanovSmooth}. 

\begin{figure}
\includegraphics[scale=0.22]{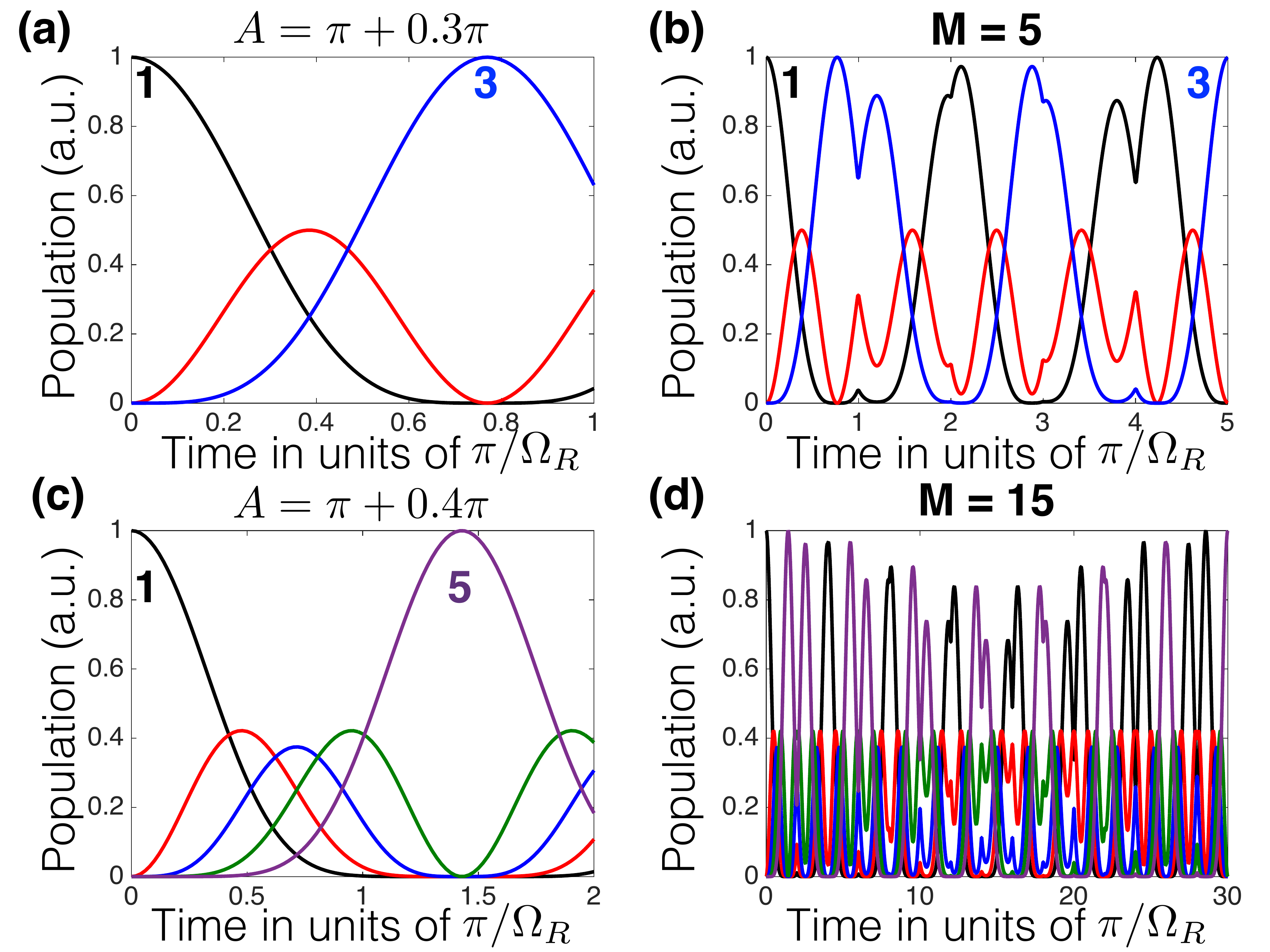}
\begin{centering}
\par\end{centering}
\caption{(Color online) \textbf{Broadband population inversion in N-level systems.}(a) Three-level system exposed to single pulse of inaccurate area $A=\pi+0.3\pi$. (b) Same system exposed to $M=5$ BB composite pulses results in full population inversion from state 1 to state 3. (c) Five-level system exposed to single pulse of inaccurate area $A=\pi+0.4\pi$. (d) Same system exposed to $M=15$ composite pulses broadband excitation results in full population inversion from state 1 to state 5. 
\label{fig:NLSBBinversions}}
\end{figure}

We now expand the composite pulse sequences for smooth temporal pulse shapes \cite{VitanovSmooth} to N-level systems, and achieve complete population inversion between levels 1 and N excited by an ill-defined ultrashort $\pi$ pulse (see Figure \ref{fig:Fig1}(b)). We find that the number of pulses $M$ dictates the achieved fidelity of the excited state. This can be shown in Figure \ref{fig:NLSBBinversions}(a) in which a three-level system is excited with a single $\pi$ pulse with an inaccuracy of $\delta=0.3\pi$. Figure \ref{fig:NLSBBinversions}(b) shows the same system, excited by $M=5$ composite pulses for broadband inversion. The system is fully steered to the excited state by a coherence time of $\tau=5\pi/\Omega_{R}$. This pulse sequence also proves useful for higher order systems, such as the five-level system shown in Figure \ref{fig:NLSBBinversions}(c) and (d). This time, a longer composite pulse sequence is necessary in order to achieve complete population inversion. We refer the reader to Appendix B for a discussion on palindromic population inversion in this case, where we present examples of various N-level systems accurately steered from level $m$ to level $m' = N-m+1$.

\begin{figure}
\includegraphics[scale=0.22]{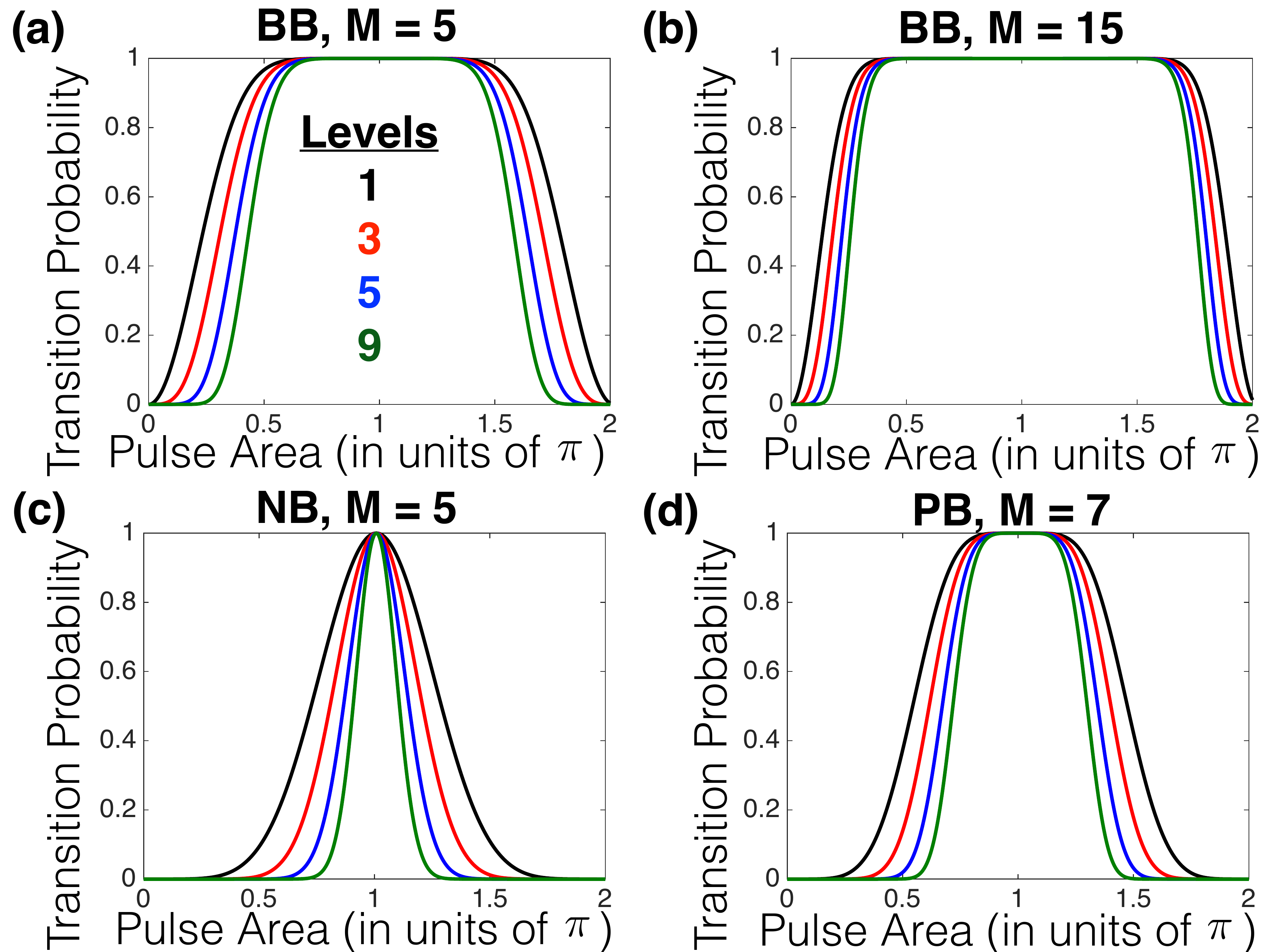}
\begin{centering}
\par\end{centering}
\caption{(Color online) \textbf{Transition probabilities for composite pulse sequences in N-level systems.} (a) Transition probabilities for $M=5$ broadband composite sequences as a function of the area of the pulses for a two-level (blue), three-level (red), five-level (yellow) and nine-level (purple) system. (b) Transition probabilities for $M=15$ broadband composite sequences as a function of the area of the pulses for the same N-level systems as (a). (c) Transition probabilities for $M=5$ narrowband composite sequences as a function of the area of the pulses for N-level systems. (d) Transition probabilities for $M=7$ passband composite sequences as a function of the area of the pulses for N-level systems.
\label{fig:NLSTransitions}}
\end{figure}

In order to present the fidelity of population inversion in N-level systems, in Figure \ref{fig:NLSTransitions} we calculate the transition probability to the desired excited level as a function of the different pulse areas in units of $\pi$. Figure \ref{fig:NLSTransitions}(a) and (b) show the transition probabilities for a broadband composite pulse sequence \cite{VitanovSmooth} comprised of $M=5$ and $M=15$ pulses respectively for a two-level (in blue), three-level (red), five-level (yellow) and nine-level (purple) system. Note that the fidelity broadens for larger values of ingredient pulses $M$ and for smaller values of levels $N$. 

Figure \ref{fig:NLSTransitions}(c) shows the transition probabilities for a narrowband composite pulse sequence \cite{VitanovSmooth} comprised of $M=5$ pulses for several N-level systems. It is noticeable that in this case, the fidelity is narrower for larger values of $N$. In Figure \ref{fig:NLSTransitions}(d), we calculated the transition probabilities as a function of the pulse area for several N-level systems, given a passband excitation of $M=7$ composite pulses \cite{VitanovSmooth}. Here, the fidelity of the passband excitation scheme narrows and sharpens for larger values of N. 

This generalization of composite ultrashort pulses in the broad framework of expanding these pulse schemes for N-level systems with SU(2) symmetry, is suitable for the coherent excitation of optical systems, in which the level spacing does not have to be equidistant. The ability to prepare quantum states with high fidelity will be very useful for quantum information processing, robust state preparation in cooling schemes, efficient coherent manipulation of atomic and molecular quantum systems. 

\section{Majorana Representation of Composite Pulses in N-level System as Points on the Bloch Sphere}

\begin{figure*}
\begin{minipage}{\columnwidth}
\includegraphics[width=3.2in]{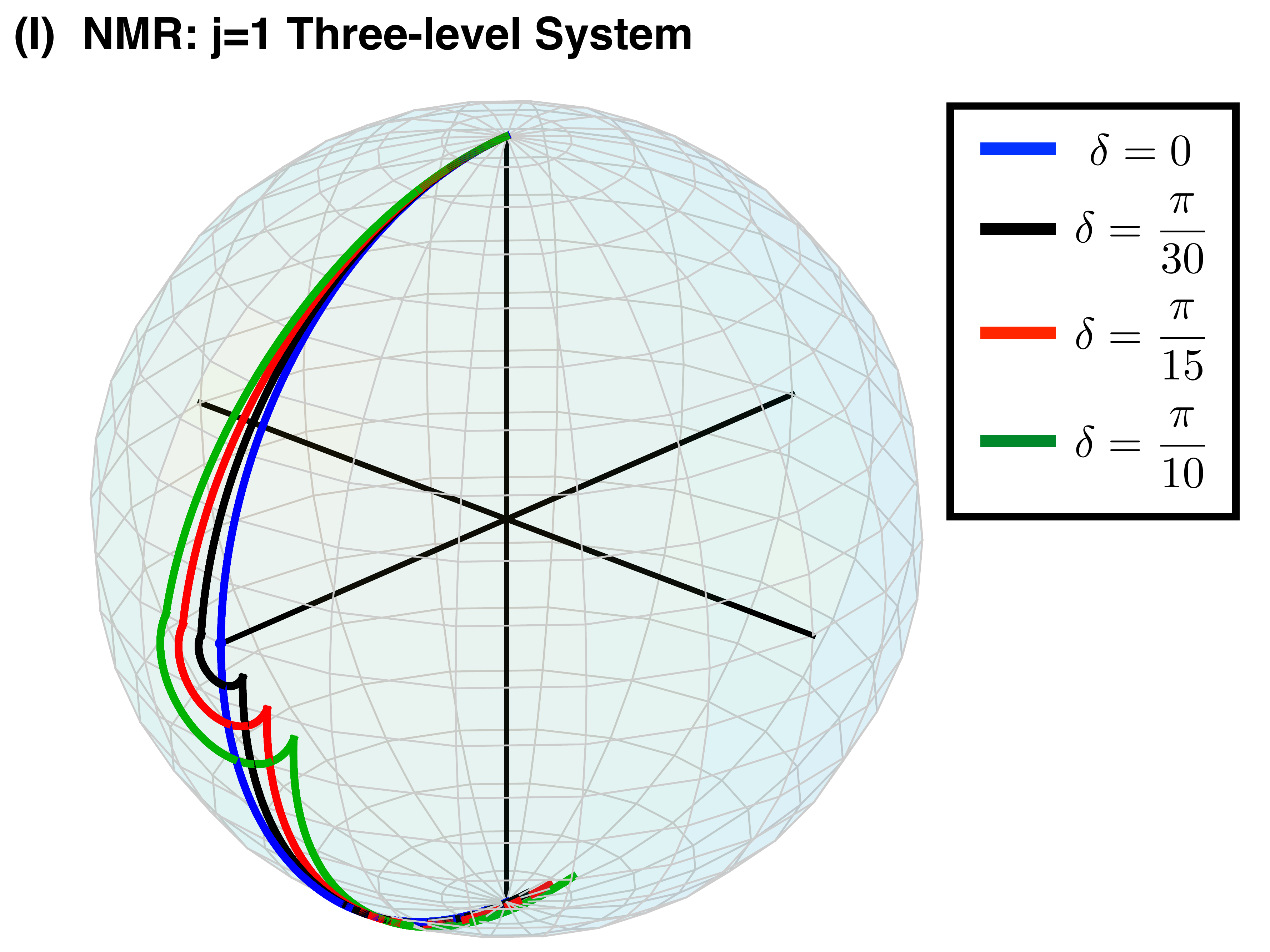}\centering
\end{minipage}
\begin{minipage}{\columnwidth}
\includegraphics[width=3.2in]{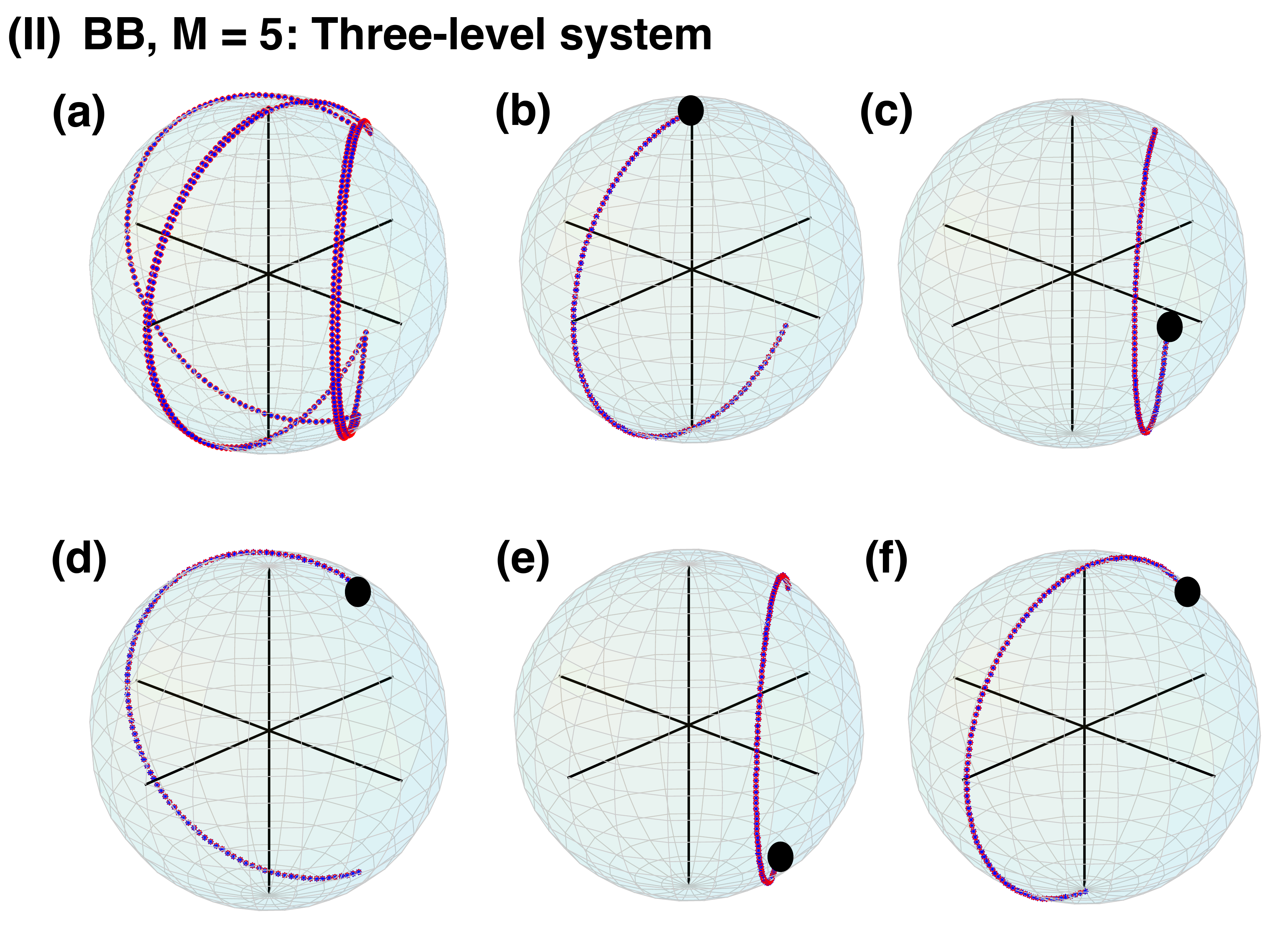}\centering
\begin{centering}
\par\end{centering}
\end{minipage}
\caption{(Color online) \textbf{Majorana representation of composite pulses in N-level systems.}(I) Trajectories of two overlapping Majorana points of a three-level spin $j=1$ system for various pulse area inaccuracies: $\delta=0$ (blue), $\delta=\pi/30$ (black), $\delta=\pi/15$ (red) and $\delta=\pi/10$ (green). (II) (a) Broadband
excitation of N=3 level system by $M=5$ composite pulses, where $\psi(t=0)=|1>$. The two Majorana points (blue and red) overlap, begin their trajectories on the North pole and evolve to the South pole according to the five trajectories shown in the following sub-figures.
(b) Overlapping trajectories of the two Majorana points (MP) during the first of five composite pulses, beginning their motion at the North pole, indicated by a black dot in the figure. (c)-(f) Trajectories of the MPs during the second, third, fourth, and fifth of five composite pulses, respectively. \label{fig:Fig5}}
\end{figure*}

The geometric interpretation of the dynamics in N-level systems is desirable in order to gain insight on these quantum states. The Majorana representation provides an elegant and compact means of portraying the evolution of an N-level system with SU(2) symmetry on a unit sphere. Here, we use this description to display the time evolution of an N-level system, as a Bloch sphere has traditionally been used to represent the dynamics of a two-level system \cite{QuantumInfo1, BornWolf, PhysRev.70.460}. 

Without loss of generality, in the Bloch sphere representation, each spin $\frac{1}{2}$ equivalent state can be written as $|\psi> = cos\frac{\theta}{2}|\downarrow>+e^{i\phi}sin\frac{\theta}{2}|\uparrow>$, where $\theta \in [0,\pi]$ and $\phi \in [0,2\pi]$. It is easily seen that the basis vectors correspond to the North and South poles of a unit sphere. Thus, the Rabi oscillations presented as solid lines in Figure \ref{fig:Fig1}(c) are conveniently described by a complete circular trajectory on a unit sphere. This geometric description was extended by E. Majorana \cite{MajoRep, RevModPhys.17.237} to describe higher order dynamics. In this representation, a spin $S$ state is described by $2S$ points on a unit sphere, each representing spin $\frac{1}{2}$ particles, coupled, so their total spin equals $S$. We employ the Majorana representation to visualize the coherent dynamics of an N-level system, driven by the composite pulse schemes introduced in the previous sections. A comprehensive derivation is available in Appendix C, with an example of composite pulse dynamics in a three-level system in Appendix D.

Imposing the Hamiltonian given by equation \ref{eq:hamiltonian} on the initial state $|1>=(1,0,0)$ results in \emph{two} Majorana points with trajectories similar to the single point on a Bloch sphere, representing the dynamics of a two-level system. In the three-level case, the two points begin on the North pole of the Majorana sphere and perform a full rotation around its $\hat{y}$ axis. For the initial state $|2>=(0,1,0)$, the trajectories of both points are complimentary, namely, one point begins its trajectory on the North pole and the other on the South pole.

The Majorana representation of the NMR composite pulse sequence for a general spin $j$ population inversion is shown in Figure \ref{fig:Fig5}(I), for the case of a three-level spin $j=1$ system, with the initial condition of $\Psi(t=0)=|1>$. This initial condition results in two overlapping Majorana points, which begin their trajectories at the North pole of the sphere. We show the time evolution of the different trajectories for the system, exposed to the NMR spin inversion pulse, for different pulse shape inaccuracies, from $\delta=0$ to $\delta=\pi/10$, in the Supplementary Material. We also show how the initial condition of $\Psi(0)=\Psi_{2}$ results in two complementary trajectories of the two Majorana points.

The various ultrashort composite pulses sequences acting on a three level system also result in informative trajectories on the Majorana sphere, as seen in Figure \ref{fig:Fig5}(II). A broadband sequence of $M=5$ composite pulses, shown in Figure \ref{fig:Fig5}(a) creates two overlapping Majorana points that quickly begin to move from their initial position on the North pole of the unit sphere, along a counter-clockwise direction. The different trajectories of the Majorana points for each of the 5 composite pulses are shown in Figures \ref{fig:Fig5}(b)-(f) (see Supplementary Material for the time evolution of these representations). Here, one sees how the three-level system is efficiently steered stepwise from its initial state at the North pole to its final state at the South pole of the Majorana sphere.

This visualization clearly shows the benefits of composite pulses in N-level systems with SU(2) symmetry. Trajectories that do not reach the South pole translate to states with lower fidelity. Composite pulse schemes enable complete population inversion, even between palindromic states, seen in the Supplementary Materials.

\section{Conclusion}
In conclusion, in this work we expanded the scheme of composite pulses to N-level systems, allowing for accurate and robust customized population transfer. We generalized the commonly used NMR spin inversion scheme for the spin $j$ case. This allows one to apply to an N-level system any composite scheme from the rich variety of sequences developed for radio frequency excitation of two-level systems. Additionally, we have shown that ultrashort composite pulses, and specifically the broadband, narrowband and passband solutions \cite{VitanovSmooth} can be utilized in multi-level systems,described within the dynamics of the irreducible SU(2) model. Our method enables the coherent control of dynamics in physical multi-level systems, without imposing the approximation to a two-level solution. We have also shown a geometric representation of the composite pulse evolution of N-level systems on the Majorana sphere. This description is advantageous, as it provides a feasible, intuitive means to grasp complex dynamics in multi-dimensional Hilbert space. We believe that this expansion of composite pulse schemes, along with such intuitive visualization will lead to new findings and possibly new solutions for coherent control of complicated higher order systems.

\appendix

\section{Coherent Dynamics of N-level Systems}
The propagator describing the evolution of an N-level system can be simpified and parametrized in terms of complex Cayley-Klein parameters\cite{Hioe:87} with a constant phase shift in the Rabi frequency:
\begin{eqnarray}
U_{\phi}(A_{t}) & = & \left(\begin{array}{cc}
a & be^{-i\phi}\\
-b^{*}e^{i\phi} & a^{*}
\end{array}\right)
\label{eq:propogatorPhi}
\end{eqnarray}
where in the case of time-independent coupling:
\begin{eqnarray}
a & = & cos\left(\frac{1}{2}\Omega_{R}t\right)-i\frac{\Delta}{\Omega_{R}}sin\left(\frac{1}{2}\Omega_{R}t\right) \nonumber\\
b & = & i\frac{\Omega_{0}}{\Omega_{R}}sin\left(\frac{1}{2}\Omega_{R}t\right)
\end{eqnarray}
and 
\begin{eqnarray}
\Omega_{R} & = & \sqrt{\Delta^{2}+\left|\Omega_{0}\right|^{2}}\\
\Omega_{0} & = & Ae^{-i\phi} \nonumber
\end{eqnarray}
In the case of an exact resonance ($\Delta=0$), the Cayley-Klein
parameters depend only on the pulse area $\hat{U}=\hat{U}(A)$.

Now, we can expand this to the N-level case. Since $\hat{H(t)}$ is given by equation \ref{eq:hamiltonian}, for different values of $t$, the vectors representing the $N$ states of an N-level system are transformed among themselves by the transformations of the SU(2) group. This means that if one finds the set of parameters $a(t)$, $b(t)$ for the two-level case, an immediate solution to the Schrodinger equation \ref{Schrodinger} is found from the $N=2j+1$ representation of the unitary group $D^{j}[a(t),b(t)]$:
\begin{eqnarray}
\psi_{m}^{(j)}(t) & = & \sum_{m'=-j}^{j}D_{mm'}^{(j)}\left[a(t),b(t)\right]C_{m'}^{(j)}(0)
\end{eqnarray}
where $m=-j,-j+1,...,j$, and the matrix elements are given as\cite{Hioe:87,Hamermesh}:
\begin{eqnarray}
D_{mm'}^{(j)} & = & \sum_{\mu}\frac{\sqrt{\left(j-m\right)!\left(j+m\right)!\left(j-m'\right)!\left(j+m'\right)!}}{p!q!r!s!} \nonumber \\
 &  & \times a^{p}a^{*q}b^{r}\left(-b^{*}\right)^{s}
\end{eqnarray}
Here, $p=j-m-\mu$, $q=j+m'-\mu$, $r=\mu$ and $s=m-m'+\mu$ and $C^{(j)}_{m'}(0)$ is the initial condition vector of the system.

Thus, in our notation, a spin $j$ system is interchangeable with an N-level system of this representation. For example, for a $j=1$ three-level system, the matrix is the following:
\begin{eqnarray}
D^{(j=1)}(a,b) & = & \left(\begin{array}{ccc}
a^{2} & \sqrt{2}ab & b^{2}\\
-\sqrt{2}ab^{*} & \left|a\right|^{2}-\left|b\right|^{2} & \sqrt{2}a^{*}b\\
-b^{*2} & -\sqrt{2}a^{*}b^{*} & a^{*2}
\end{array}\right)
\end{eqnarray}

\section{Palindromic Population Inversion in N-level Systems via Broadband Ultrashort Composite Pulses}
\begin{figure}
\includegraphics[scale=0.22]{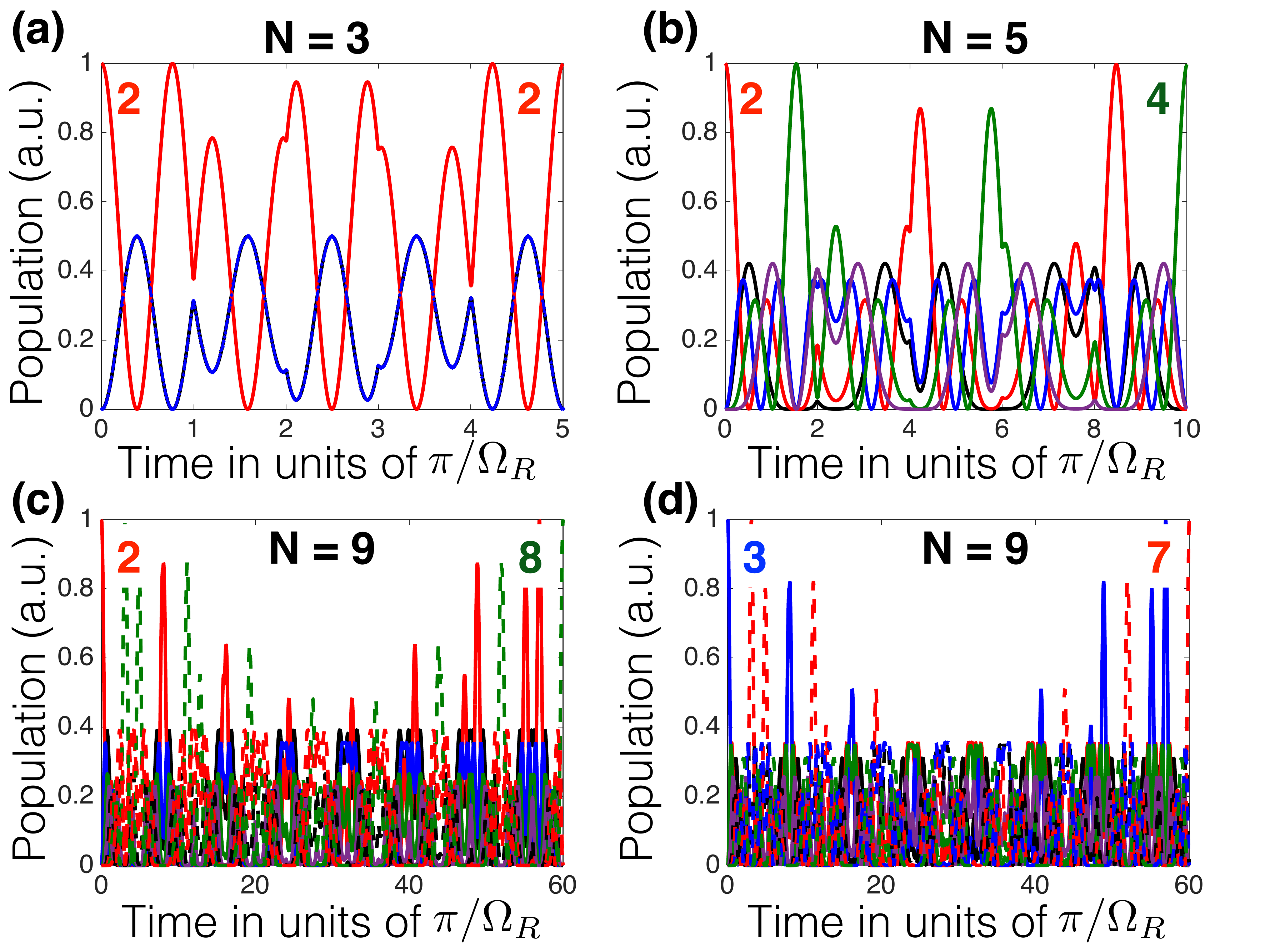}
\begin{centering}
\par\end{centering}
\caption{(Color online) \textbf{Palindromic population inversion in N-level systems via broadband composite pulses.} (a) Intensity evolution of a three-level system with the initial condition $\psi(0)=\psi_{2}$ excited by $M=5$ broadband composite pulses. (b) Five-level system excited by $M=5$ broadband composite pulses, with the initial condition $\psi(0)=\psi_{2}$ . (c) Nine-level system excited by $M=15$ broadband composite pulses, with the initial condition $\psi(0)=\psi_{2}$ . (d) Same system and excitation as (c), with the initial condition $\psi(0)=\psi_{3}$.
\label{fig:InitialConditions}}
\end{figure}

Taking advantage of the unique property of palindromic excitation of N-level systems, we calculated the evolution
of several such exposed to a broadband composite pulse scheme \cite{VitanovSmooth} with different initial conditions. Figure \ref{fig:InitialConditions}(a) shows a three-level system excited by $M=5$ such pulses, with the initial condition of $\psi(0)=\psi_{2}$. The system remains in this state, as it is its own palindromic counterpart. Figure \ref{fig:InitialConditions}(b) is a five-level system excited by the same broadband composite pulse sequence, with the initial condition of $\psi(0)=\psi_{2}$. This time, the system evolves into the final state $\psi_{4}$. Figures \ref{fig:InitialConditions}(c) and (d) present the evolution of a nine-level system excited by a broadband composite sequence of $M=15$ pulses, with the initial conditions $\psi_{2}$ and $\psi_{3}$, which evolve into the final states $\psi_{8}$ and $\psi_{7}$ respectively.

\section{The Majorana Representation}
The basis for the spin states of a spin-S particle is given as 
\[
|S,M>,\:M=-S,...,+S
\]
thus an arbitrary pure spin state can be written as $|\xi>=\sum_{M=-S}^{S}\xi_{M}|S,M>$,
where $\xi_{M}$ is a complex number for each M. We define the spin $S$ state as a  symmetrized tensor product of spin $\frac{1}{2}$ states, $|\tilde{\xi}>=\frac{A}{\left(2S\right)!}\sum_{P}\otimes_{k=1}^{2S}\phi_{k}$, with components given by $\phi_{k}=\alpha_{k}|\uparrow>_{k}^{M}+\beta_{k}|\downarrow>_{k}^{M}$,
where $k=1,..,2S$, and where $P$ are the permutations of the signs $\uparrow$ and $\downarrow$.

To obtain the spin $\frac{1}{2}$ vectors, if $|\tilde{\xi}>$ is
known, one must define the Majorana polynomial \cite{MajoRep}

\begin{eqnarray}
P_{Majo}\left(|\tilde{\xi}>;x\right) & = & \sum_{M=-S}^{S}\left(\begin{array}{c}
2S\\
S+M
\end{array}\right)^{1/2}\tilde{\xi}_{M}x^{S+M}\label{eq:majo1}
\end{eqnarray}

Here,we denote a spin $S$ system by $2S$ decimal numbers, such that:
\begin{eqnarray}
|i>_{d} & = & \sum_{j=0}^{2S}|S,j-S>
\end{eqnarray}
where $i=0,1,2,...,2S$. Given a state vector of a spin $S$ system
$\phi=\sum_{i=0}^{2S}C_{i}|i>_{d}$, where $C_{i}$ is the weight of each of the $2S$ states $|i>_{d}$, the corresponding Majorana polynomial
becomes:

\begin{eqnarray}
P_{Majo}\left(\phi;x\right) & = & \sum_{i=0}^{2S}\left(\begin{array}{c}
2S\\
i
\end{array}\right)^{1/2}C_{i}x^{i}\label{eq:MajoPoly}
\end{eqnarray}
Using the roots of the Majorana polynomial, every spin $S$ state
can be mapped as $2S$ points on the Bloch sphere \cite{NQubitStates}, namely
\begin{equation}
x_{k}=tan\frac{\theta_{k}}{2}e^{i\phi_{k}}\label{eq:rootsMajo}
\end{equation}
where $\theta_{k},\phi_{k}$ are the angular coordinates on a sphere.
In equation \ref{eq:rootsMajo}, one can see that $x_{k}=-\beta_{k}/\alpha_{k}$,
thus $\alpha_{k}=0$ corresponds to $\theta_{k}=\pi$. 

It can be shown\cite{NQubitStates} that if the Majorana polynomial defined as in equation
\ref{eq:majo1} is a rotation of the spin vector $|\xi>$ by a spin
rotation matrix $D^{(S)}\left(\alpha,\beta,\gamma\right)$, this corresponds
to rotating the point configuration of $|\xi>$ by $R\left(\alpha,\beta,\gamma\right)\in SO\left(3\right)$,
parametrized by the Euler angles.

Therefore, applying the hamiltonian given by equation \ref{eq:hamiltonian} on an N-level system acts as a rotation of all the Majorana points on the sphere. Thus, the trajectory of these points are interpreted as the evolution of the multi-level system dynamics.

\section{Example of Majorana Representation for Composite Pulse Dynamics in a Three-level System}
In the following we will demonstrate this visualization procedure for composite pulse dynamics in a three-level system. Consider a three level spin $1$ system comprised of three states:
$|1>\equiv|1\, ,-1>\equiv(1,0,0)$ , $|2>\equiv|1\, ,0>\equiv(0,1,0)$ and $|3>\equiv|1\, ,1>\equiv(0,0,1)$, with dynamics described by
the Hamiltonian given by equation \ref{eq:hamiltonian}. This time: 
\begin{eqnarray}
\hat{H} & = & \frac{\pi}{\sqrt{2}}\left(\begin{array}{ccc}
0 & 1 & 0\\
1 & 0 & 1\\
0 & 1 & 0
\end{array}\right)
\label{3LSHam}
\end{eqnarray}

The eigenvalues and eigenvectors of this Hamiltonian $AV=VD$ are
\begin{eqnarray*}
D & = & \pi\left(\begin{array}{ccc}
-1 & 0 & 0\\
0 & 0 & 0\\
0 & 0 & 1
\end{array}\right)\;V=\frac{1}{2}\left(\begin{array}{ccc}
1 & -\sqrt{2} & 1\\
-\sqrt{2} & 0 & \sqrt{2}\\
1 & \sqrt{2} & 1
\end{array}\right)
\end{eqnarray*}

We use the Majorana representation to describe these eigenvectors
in terms of those of a two level system, completely represented by
the basis of two eigenstates $|\uparrow>$ and $|\downarrow>$.
One could define the symmetric (S) and antisymmetic (A) states for
each spin $1/2$:

\begin{eqnarray}
|S & >_{i} & =\frac{|\uparrow>_{i}+|\downarrow>_{i}}{\sqrt{2}}
\end{eqnarray}
\begin{eqnarray*}
|A & >_{i} & =\frac{|\uparrow>_{i}-|\downarrow>_{i}}{\sqrt{2}}
\end{eqnarray*}

where for a spin $S$ system, $i=1,...,2S$ and in the case where $S=1$, $i=1,2$.

Recalling that each of the state vector of a two-level system on the Bloch sphere can be described by $|\psi>=cos\frac{\theta}{2}|\uparrow>+e^{i\phi}sin\frac{\theta}{2}|\downarrow>$, we now extend this to the N-level case.
By calculating the roots of the Majorana polynomial in equation \ref{eq:MajoPoly}
for the different eigenvectors $\overrightarrow{V}$, we find the
values of $\left(\theta_{i},\phi_{i}\right)$ for the $2S=2$ points
on the Majorana sphere, thus enabling a convenient representation
of the spin 1 system in the basis of two spin $1/2$ states:

\begin{eqnarray}
\overrightarrow{v_{\alpha}} & = & |S>_{1}+|S>_{2} \nonumber \\
\overrightarrow{v_{\beta}} & = & |S>_{1}+|A>_{2} \nonumber\\
\overrightarrow{v_{\gamma}} & = & |A>_{1}+|A>_{2}
\end{eqnarray}

Now, we describe the different levels of a spin 1 system in terms
of these eigenvectors:

\begin{eqnarray*}
|1> & = & \frac{1}{2}\left(\overrightarrow{v_{\alpha}}+\overrightarrow{v_{\gamma}}-\sqrt{2}\overrightarrow{v_{\beta}}\right)
\end{eqnarray*}
\begin{eqnarray*}
|2> & = & \sqrt{2}\left(\overrightarrow{v_{\gamma}}-\overrightarrow{v_{\alpha}}\right)
\end{eqnarray*}
\begin{eqnarray}
|3> & = & \frac{1}{2}\left(\overrightarrow{v_{\alpha}}+\overrightarrow{v_{\gamma}}+\sqrt{2}\overrightarrow{v_{\beta}}\right)
\end{eqnarray}

\bibliography{comppulses}

\end{document}